\documentclass{aa501}
\usepackage{graphicx,natbib}
\bibpunct{(}{)}{,}{a}{}{,}
\begin{document}
\title{Flare Generated Acoustic Oscillations in Solar 
and Stellar Coronal Loops}
\author{D. Tsiklauri \inst{1} \and V.M. Nakariakov \inst{2} 
\and T.D. Arber \inst{2} \and M.J. Aschwanden \inst{3}}
\offprints{David Tsiklauri, \\ \email{D.Tsiklauri@salford.ac.uk}}
\institute{Joule Physics Laboratory,
School of Computing, Science \& Engineering,
University of Salford, Salford, M5 4WT, England, UK 
 \and  Physics Department, University of Warwick, Coventry,
   CV4 7AL, England, UK \and Lockheed Martin, Advanced Technology Center
Solar \& Astrophysics Laboratory, Dept. L9-41, Bldg.252
3251 Hanover Street, Palo Alto, CA 94304, USA}
\date{Received ???  2004 / Accepted ??? 2004}

\abstract{Low-frequency longitudinal oscillations of a 
flaring coronal loop are
 studied numerically. 
 In the recent work of Nakariakov et al. 
 A\&A, 414, L25-L28 (2004) it has been shown that 
the time dependences of density and velocity
in a flaring loop contain well-pronounced quasi-harmonic 
oscillations
associated with a 2nd harmonic of a standing slow magnetoacoustic wave.
In this work we investigate physical nature of these oscillations
in greater detail, namely, we study their spectrum (using periodogram technique)
and how does heat positioning affects the mode excitation.
We found that excitation of such oscillations
practically independent of the positioning of the
heat deposition in the loop. Because of the change of the
background temperature and density, the phase shift between the
density and velocity perturbations is not exactly equal to the
quarter of the period, it varies along the loop and
is time dependent, especially in the case of one footpoint (asymmetric)
heating.
\keywords{Sun: flares -- Sun: oscillations -- Sun: Corona -- 
Stars: flare -- Stars: oscillations -- Stars: coronae } }
\titlerunning{Flare Generated Acoustic Oscillations ...}
\authorrunning{Tsiklauri et al.}
\maketitle

\section{Introduction}
Magnetohydrodynamic (MHD) coronal seismology is the main
reason for 
studying waves  in the solar corona. Also such
studies are important in connection
with coronal heating and solar wind acceleration problems. 
Observational evidence in the EUV coronal emission 
of coronal waves and oscillations is
numerous (e.g. \citep{o99,ow02}). 
Radio
band observations also demonstrate various kinds 
of oscillations (e.g., the
quasi-periodic pulsations, or QPP, see \citep{a87} for a
review), usually with the periods from a few seconds to tens 
of seconds.  Also,
decimeter and microwave observations  show much
longer periodicities, often in association with a flare. For
example, \citep{wx00} observed QPP with the periods of about
50~s at 1.42 and 2~GHz (in association with an M4.4 X-ray flare).
Similar periodicities have been observed in the X-ray band  (e.g.,
\citep{m97,terekhov02}) and in 
the white-light emission associated with the stellar
flaring loops \citep{mathio}.
A possible interpretation of these medium period QPPs may be
interpreted in terms of kink or torsional modes \citep{zs89}. 
In our previous, preliminary  study 
 \citep{nt04}, we have outlined an alternative, {\it simpler, thus more
aesthetically attractive}, mechanism for generation
of long-period QPPs. In this study we {\it demonstrate
in detail,} 
that in a coronal loop an
impulsive (time-transient) 
energy release efficiently generates the second spatial
harmonics of a slow magnetoacoustic mode. 
In particular, in the present study we study 
the spectrum (using periodogram technique) of these oscillations
and how does heat positioning affects the mode excitation.

\section{Numerical Results}

The model that we use to describe plasma dynamics in a coronal loop
is outlined in \citep{nt04}. Here we just add that,
when solving
numerically 1D radiative hydrodynamic equations (infinite magnetic
field approximation), and using a 1D version of Lagrangian Re-map
code (Arber et al. 2001) with the radiative loss limiters, 
the radiative loss function was specified as
$$
L_r(T)=n_e^2 
\left\{ \matrix{10^{-26.60} \; T^{1/2} & T>10^{7.6}  \cr
10^{-17.73} \; T^{-2/3} & 10^{6.3} < T < 10^{7.6}  \cr
10^{-21.94}  & 10^{5.8} <T < 10^{6.3}  \cr  
10^{-10.40} \; T^{-2} & 10^{5.4} <T < 10^{5.8}  \cr   
10^{-21.2}  & 10^{4.9} <T < 10^{5.4}  \cr      
10^{-31} \; T^{2} & 10^{4.6} < T < 10^{4.9}  \cr     
10^{-21.85}  &$ $10^{4.3} < T < 10^{4.6}  \cr     
10^{-48.31} \; T^{6.15} & 10^{3.9} < T < 10^{4.3}  \cr      
10^{-69.90}\; T^{11.7} & 10^{3.6}<T < 10^{3.9} \cr} \right.
\label{1}
$$
which is \citep{rtv78} law extended to a wider temperature range
\citep{pe82,p82}.

We have used the same heating function as in \citep{nt04}.
The choice  of the temporal
part of the heating function is such that at all times
there is a small background heating present (either at footpoints or
the loop apex) which ensures that in the absence of flare heating
(when $\alpha$, that determines flare heating amplitude, is zero)
the average loop temperature
stays at 1 MK.
For easy comparison between that apex and footpoint heating
cases, with fix, $Q_p$, flare heating amplitude at a given different 
value in each case (which ensures that with the flare heating on
when $\alpha=1$
the average loop temperature
peaks at about the observed value of 30 MK in {\it both cases}).

In all our numerical runs, presented here,  $1/(2 \sigma_s^2)$
was fixed at 0.01 Mm$^{-2}$, which gives the heat deposition length scale, 
$\sigma_s=7$ Mm. This is a typical value determined from
the observations  \citep{abk02}. 
The flare peak time was fixed in our numerical
simulations at 2200 s. 
The duration of flare, $\sigma_t$, was fixed at 333 s.
The time step of data visualization
(which in fact is much larger than the actual time step, 0.034 s, in
the numerical code) was chosen to be 0.5 s.

\subsection{Case of Apex (Symmetric) Heating}

In this subsection we
complete analysis started in \citep{nt04}, namely for the same 
numerical run we study spectrum of the detected oscillations
at different spatial points.

\begin{figure}[]
\resizebox{\hsize}{!}{\includegraphics{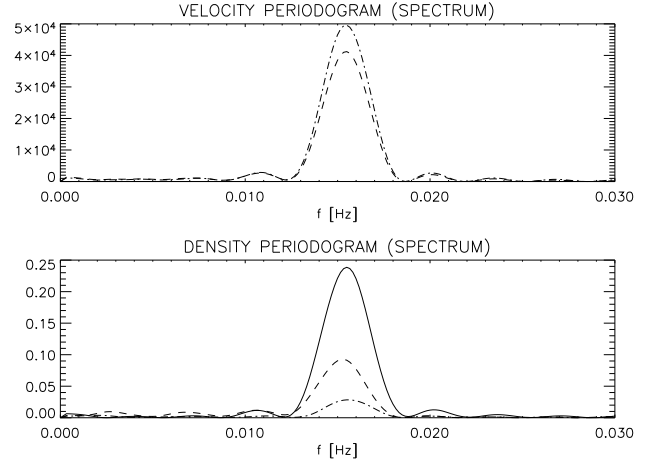}} 
\caption{Case of apex (symmetric) heating: Periodogram
(spectrum) of the velocity and density oscillatory
component times series outputted in the following three points: 
loop apex
 (solid curve), 
$1/4$ (dash-dotted curve) and $1/6$ (dashed curve)
of the effective loop length (48 Mm), i.e. at $s=0,-12,-16$ Mm.} 
\end{figure}

\begin{figure}[]
\resizebox{\hsize}{!}{\includegraphics{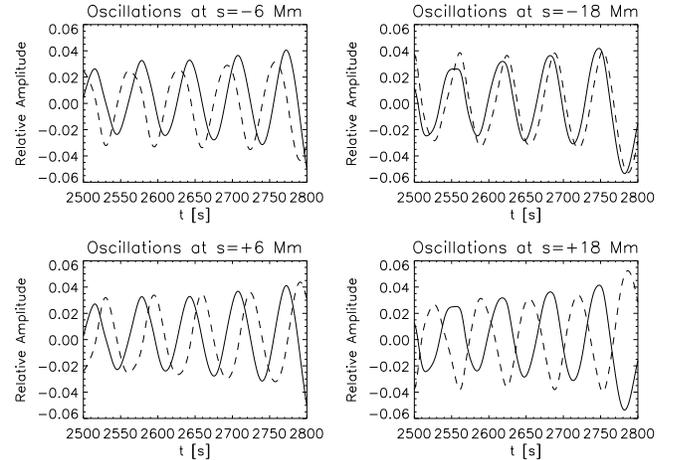}} 
\caption{Case of apex (symmetric) heating:
oscillatory components of time series outputted at 
$\pm 6$ Mm and $\pm 18$ Mm in   
the time interval of 2500-2800 s.
The solid curve shows
plasma number density in units of $10^{11}$ cm$^{-3}$. 
The dashed curve shows velocity  normalized to 400 km s$^{-1}$.} 
\end{figure}

As it was pointed out in 
\citep{nt04},
the most interesting fact is that we see clear quasi-periodic
oscillations, especially in the second stage (peak of the flare)
for time interval $t=2500-2800$ s (cf. Fig.~1 in \citep{nt04}).
In fact, such oscillations are frequently seen 
during the solar flares observed in X-rays, 8-20 keV, 
(e.g. \citet{terekhov02}) as well as stellar flares observed
in white-light (e.g. \citet{mathio}).

Before we start discussion of the
physical nature of these oscillations,
first let us recall for completeness what simple 1D analytic theory
of standing sound waves tells us: If we use 1D, linearized, hydrodynamic 
equations with constant unperturbed (zero order) background variables,
the solutions for density, $\rho$ and velocity, $V_x$, can be
easily written as
\begin{equation}
V_x(s,t)=A \cos \left( {{n \pi C_s}\over{L}} t\right)
\sin \left( {{n \pi}\over{L}} s\right),
\end{equation}
\begin{equation}
\rho(s,t)=-{{A \rho_0}\over{C_s}} \sin \left( {{n \pi C_s}\over{L}} t\right)
\cos \left( {{n \pi}\over{L}} s\right),
\end{equation}
where $C_s$ is a speed of sound, $A$ is wave amplitude, $L$ is
loop length, $n=1,2,3,...$ is a number of harmonic, and $s$ is a distance
along the loop.
Note that (relative) phase shift between $V_x$ and $\rho$
is $\Delta P/ P=-(\pi/2)/(2 \pi)=-1/4$, where $P$ is standing wave
period.

In Fig.~1 we present  periodogram
(spectrum) of the velocity and density oscillatory
component times series outputted in the following three points: 
loop apex, 
$1/4$ and $1/6$ of the effective loop length (48 Mm), 
i.e. at $s=0,-12,-16$ Mm.
The choice first two points is obvious because we wanted to test whether
simple analytic solution for 1D standing sound waves (see below)
is at all relevant in this case. The third point ($1/6$) was chosen
arbitrarily (any spatial point along the loop where density and
velocity of the standing waves does not have a node would be
equally acceptable).
We gather from the graph that as expected for 2nd spatial
harmonic of a standing sound wave in the velocity periodogram
there are two clearly defined peaks and the largest peak
corresponds to $1/4$ of the effective loop length, while
smaller peak corresponds to the  $1/6$. Note that at the
loop apex periodogram gives $0$ (solid line is too close to zero
to be seen in the plot). While density periodogram is exactly
opposite to the velocity one. Largest peak
corresponds to the loop apex, while $1/6$ of the effective loop length
corresponds to a smaller peak, and $1/4$ is quite small
(close to zero). 
Location of the peaks are at about 0.0155 Hz i.e. period of
the oscillation is 64 s. Period of a 2nd spatial
harmonic of a standing sound wave should be 
\begin{equation}
P=L/C_s=L/(1.52 \times 10^5 \sqrt{T}),
\end{equation}
 where
 $T$ is plasma temperature measured in MK.
If we substitute effective loop length $L=48$ Mm (see Fig.~2 in 
\citep{nt04}) and
an average temperature of 25 MK (see top panel in Fig.~1 in 
\citep{nt04} in the 
range of 2500-2800 s -- quasi periodic oscillations time interval we study)
we obtain 63 s, which is quite close to the result of our numerical
simulation. In fact, such a close coincidence is surprising
bearing in mind that the theory does not take into
account variation of background density and velocity in time, while
we see from Fig.~1 in \citep{nt04} that even within considered short interval
during the flare 2500-2800 s all physical quantities
vary in time significantly.
To close our investigation of the physical nature of
the oscillations, we study the phase shift between 
velocity and density oscillations, and
compare our simulation results with the 1D analytic theory.
In Fig.~2 we plot 
oscillatory components of time series outputted at 
$\pm 6$ and $\pm 18$ Mm  
of plasma number density in units of $10^{11}$ cm$^{-3}$
and velocity, normalized to 400 km s$^{-1}$ 
(note that sound speed at
$25$ MK is 760 km s$^{-1}$ which roughly is within the
$t=2500-2800$ s). 
These points where selected so that one
symmetric (with respect to the apex) pair ($\pm 6$ Mm)
is close to the apex, while another pair
($\pm 18$ Mm) is closer to footpoints.
We gather  from this graph: (A) clear quasi periodic
oscillations are present; (B) that they are shifted with respect 
to each other in time;
(C) in the case of the pair close to the apex ($\pm 6$ Mm)
the phase shift is quite close to the one
predicted by the 1D analytic theory;
(D) in the case of the pair close to the footpoints ($\pm 18$ Mm)
the phase shift is somewhat different from the one
predicted by the 1D analytic theory.
In the latter case the discrepancy can be attributed
to the presence flows close to the footpoints.
Main reason for the overall deviation is due to fact that analytic
theory does not take into
account variation of background density and velocity in time,
and the fact that density gradients in the
transition region are not providing perfect
reflecting boundary conditions for the
formation of standing sound waves.

Yet another interesting result is that even with wide variation
of the parameter space of the flare, its duration, peak average
temperature, etc. we always obtained 
dominant 2nd spatial harmonic of a standing sound wave with 
some small admixture of 4th and sometimes 6th harmonic.
Our initial guess was that this is due to the symmetric excitation
of these oscillations (recall that we use apex heat deposition).
In order to investigate this the issue of excitation further
we decided to break symmetry and put heating source at one footpoint,
hoping to see excitation of odd harmonics 1st, 3rd, etc.

\subsection{Case of Single Footpoint (Asymmetric) Heating}

In the case of single footpoint heating we fix $s_0= 30$ Mm in Eq.(1) in \citep{nt04}, i.e.
(spatial) peak of the heating is chosen to be at the bottom 
of the transition region (top of chromosphere).
Initially we run our numerical code without flare heating,
i.e. we put $\alpha=0$ (in this manner we turn off flare 
heating). $E_0=0.02$ erg cm$^{-3}$ s$^{-1}$, 
was chosen such that in the steady (non-flaring) case
the  average loop temperature stays at
about 1 MK.
Then, we run our numerical code with flare heating,
i.e. we put $\alpha=1$, and fix $Q_p$ at $1 \times 10^4$,
so that it yields peak average temperature of about 30 MK.
The results are presented in Fig.~3. 
We gather from the graph that during the
flare apex temperature peaks at 38.38 MK while 
the number density at the apex at $3.11 \times 10^{11}$ cm$^{-3}$.
In this case, as opposed to the case of symmetric (apex)
heating, velocity dynamics is quite different.
Namely, since the symmetry of heating is broken, there is non zero
net flow through the apex at all times.
However, as in the symmetric heating case, 
we again see quasi-periodic
oscillations superimposed on the dynamics of all physical
quantities (cf time interval of 
$t=2400-2700$ in Fig.~3). In what follows we do analysis of their
physical nature as in the previous subsection.

In Fig.~4 we present time-distance
plots of
velocity  and density in the time 
interval 2400-2700s, where the quasi-periodic
oscillations are most clearly seen.
Here we again subtracted the slow varying with respect to
oscillation period background.
The picture is quite different from the case of apex (symmetric)
heating (compare it with Fig.~2 in \citep{nt04}). 
It is looks more complex because in this case the
node in the velocity (at the apex) 
moves back and forth  periodically along the apex,
and for larger times ($t>2550$ s) stronger flows
(darker bands with at a slope) are now present.
However, physical nature of the oscillations remains 
mainly the same
-- 2nd spatial harmonic of a sound wave 
(though with oscillating node
along the apex).

To investigate this further we plot in Fig.~5 periodogram
(spectrum) of the velocity and density oscillatory
component times series outputted in the following three points: 
loop apex, $1/6$ and $1/4$ of the effective loop length.
We gather from the graph that the  periodogram
(spectrum) is more complex than in the case of apex (symmetric)
heating. Namely,  in the velocity periodogram
at the apex we see a peak with frequency of the oscillation
different (higher) than that of 2nd spatial harmonic of
a standing sound wave. Actually that is the frequency with
which node of the velocity oscillates (see discussion in the 
previous paragraph). It has nothing to do with the standing
mode, and it is rather dictated by the excitation conditions
in the dynamics resonator, which the considered loop obviously
is. Let us analyze now how this periodogram compares
with the 1D analytic theory. Peak in the periodogram
corresponding to $1/6$ of the effective loop length
(dashed line) corresponds to frequency of about 0.017 Hz, i.e.
period of the oscillation is 59 s.
Again, the period of a 2nd spatial
harmonic of a standing sound wave should be 
calculated from Eq.~(3).
If we substitute effective loop length $L=48$ Mm (see Fig.~4) and
an average temperature of 26 MK (see top panel in Fig.~3 in the 
range of 2400-2700 s -- quasi periodic oscillations time interval we study)
we obtain 62 s, which is quite close to the result of our numerical
simulation (59 s).

Next, we study the phase shift between velocity and density
oscillations, and
compare our simulation results with the 1D analytic theory.
In Fig.~6 we produce a plot similar to Fig.~2,
but for the case of asymmetric heating.
The deviation which is greater than in the case apex (symmetric)
heating can again be attributed to the over-simplification of the
1D analytic theory, which does not take into account time variation
of the background physical quantities and 
imperfection of the reflecting boundary conditions
(see above).
And, more importantly, in the asymmetric case strong
flows are present throughout of the flare simulation time,
thus, if, say, linear time dependence would be assumed,
which is relatively good within considered short interval
during the flare 2400-2700 s, the Eqs(1)-(2) would
be modified such that phase shift would vary 
secularly in time, which is similar to what we see in Fig.~6.

Yet another interesting observation comes from the
following argument: in a steady 1D case analytic theory
predicts that the phase shift between the density
and velocity should be (A) zero for for a {\it propagating} 
acoustic wave and (B) quarter of a period for the
{\it standing} acoustic wave.
Since in the asymmetric case strong flows are present,
we see less phase shift between the velocity and
density in Fig.~6 as one would expect.

Thus, the results of the present study
provide further {\it more definitive} than in \citep{nt04}
proof  that these oscillations indeed are
2nd spatial harmonic of a standing sound wave. However, 
present work also revealed that in the case
of single footpoint (asymmetric) heating the physical nature
of the oscillations is more complex as the node in the velocity
oscillates along the apex and net flows are present.

\begin{figure}[]
\resizebox{\hsize}{!}{\includegraphics{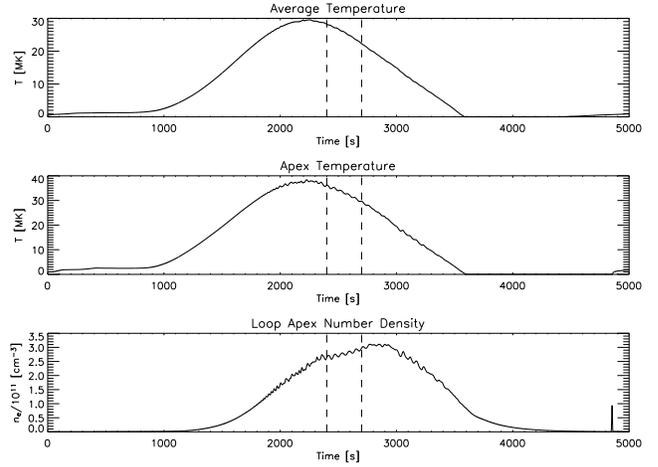}} 
\caption{Case of single footpoint (asymmetric) heating: 
Average temperature, temperature at apex, and 
 number density at apex as
a function of time.}
\end{figure}

\begin{figure}[]
\resizebox{\hsize}{!}{\includegraphics{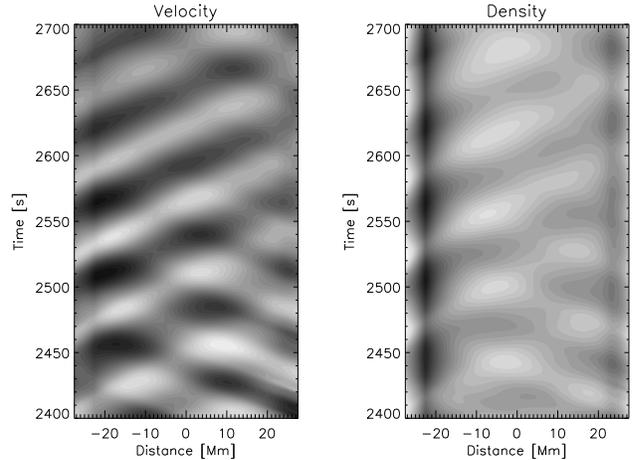}} 
\caption{Time-distance plots of the velocity and 
density oscillatory components in the
time interval of 2400-2700s for the case of single footpoint (asymmetric) 
heating.} 
\end{figure}

\begin{figure}[]
\resizebox{\hsize}{!}{\includegraphics{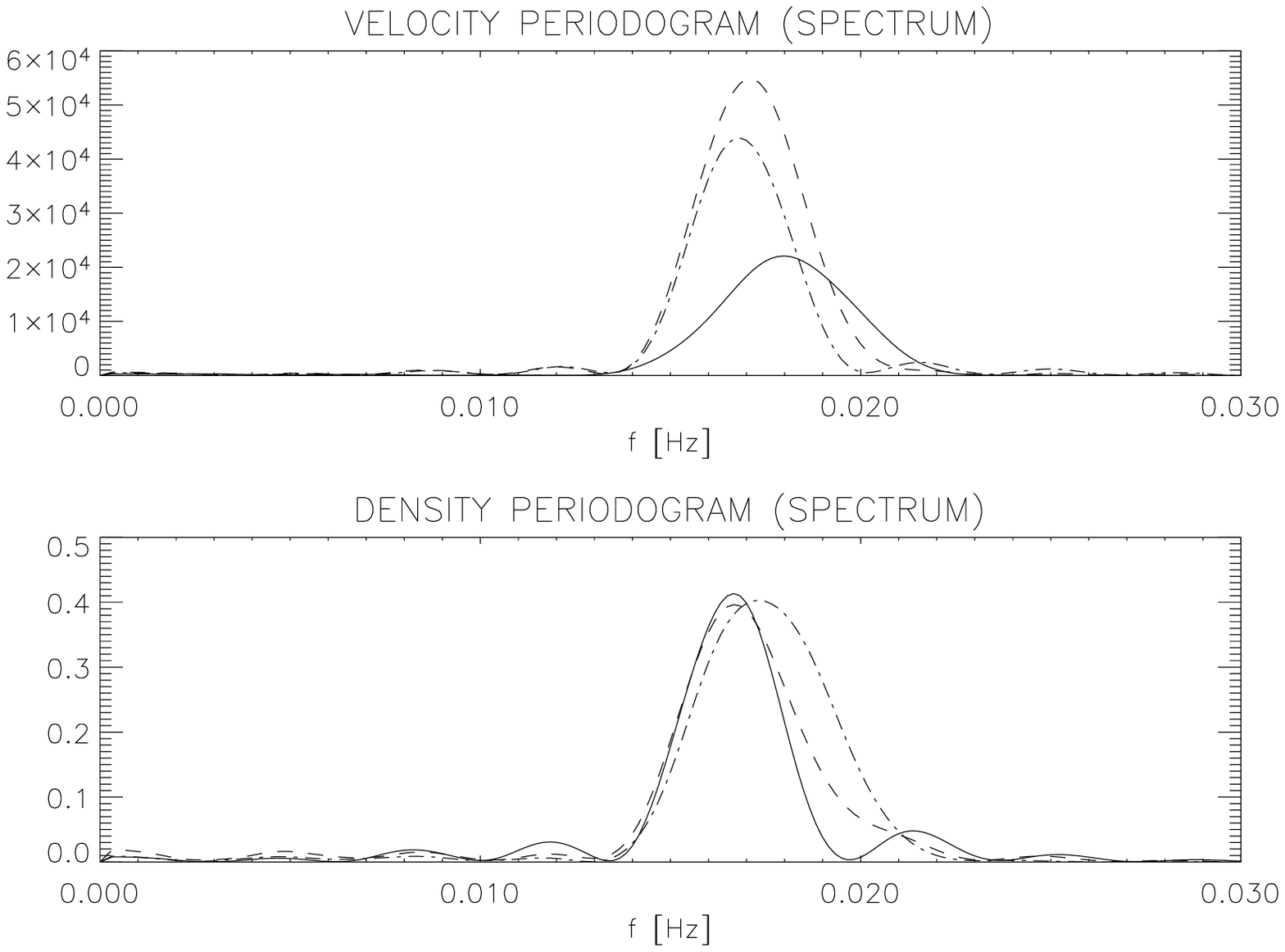}} 
\caption{As is Fig.~1 but for the
case of single footpoint (asymmetric) heating.
Time interval here is 2400-2700s.} 
\end{figure}

\begin{figure}[]
\resizebox{\hsize}{!}{\includegraphics{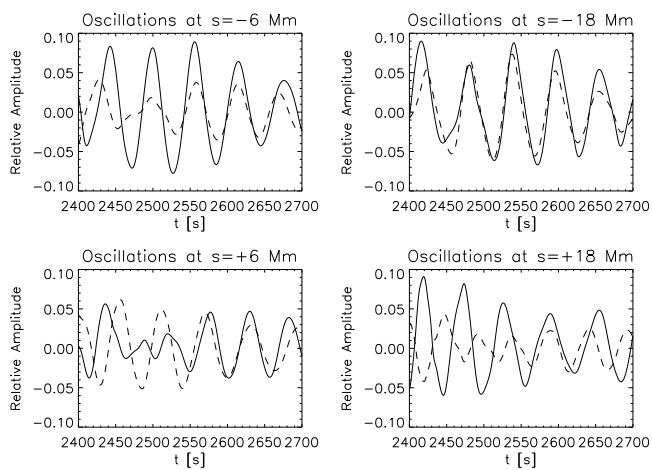}} 
\caption{As in Fig.~2, but for the case of single footpoint (asymmetric) 
heating. Time interval here is 2400-2700s.} 
\end{figure}

\section{Conclusions}

Initially we have used 1D radiative hydrodynamics 
loop model which incorporates
the effects of gravitational
stratification,
heat conduction, radiative losses, added external 
heat input, presence of Helium, hydrodynamic 
non-linearity, and bulk Braginskii viscosity
to simulate flares \citep{t04}. 
As a byproduct of that study, 
in practically all our numerical runs we have detected
quasi periodic oscillations in all physical quantities \citep{nt04}.
In fact, such oscillations are frequently seen 
during the solar flares observed in X-rays, 8-20 keV 
(e.g. \citet{terekhov02}) as well as stellar flares observed
in white-light (e.g. \citet{mathio}).
Our present analysis shows, {\it in detail}, that quasi periodic
oscillations seen in our numerical simulations
bear many similar features as the observed ones.
In summary \citep{nt04} and the present study established
 the following features:
\begin{itemize}
\item We show that the time dependences of density and temperature
 in a flaring loop contain well-pronounced quasi-harmonic oscillations
 associated with standing slow magnetoacoustic modes of the loop.
\item
For {\it a wide range of physical parameters}, 
the dominant mode is the second spatial harmonic, with a velocity
 oscillation node and the density oscillation maximum at the loop apex.
{\it This result is practically independent of the positioning of the
 heat deposition in the loop}. 
\item 
 Because of the change of the
 background temperature and density, 
 and the fact that 
density gradients in the
transition region are not providing perfect
reflecting boundary conditions for the
formation of standing sound waves,
 the phase shift between the
 density and velocity perturbations is not exactly equal to the
 quarter of the period.
\item 
 We conclude that the oscillations in the white light,
 the radio and X-ray light curves observed during
 solar and stellar flares may be produced by the slow standing mode,
with the period determined by the loop temperature and length. 
\item
For a typical
 solar flaring loop the period of oscillations is shown to be about
 a few minutes, while amplitudes are typically few percent.
\end{itemize}

The novelties brought about by this study are that by studying the spectrum 
and phase shift of these oscillations we provide more definite proof
that these oscillations are indeed 2nd harmonic of a standing sound wave,
and that the single footpoint (asymmetric) heat positioning
still produces 2nd spatial harmonic, though it is more complex than the
apex (symmetric) heating (due to the presence of flows). 

\begin{acknowledgements}
This research was supported in part by PPARC, UK.
Numerical calculations of this work were
performed using the PPARC funded Compaq MHD Cluster at St Andrews
and Astro-Sun cluster at Warwick.
\end{acknowledgements}

\end{document}